\documentclass[12pt,preprint]{aastex}
\usepackage{amsmath,longtable,natbib}
\usepackage{xcolor}
\usepackage{float}
\usepackage[caption = false]{subfig}
\usepackage{hyperref}
\usepackage{graphicx}
\shorttitle{J1227$-$4853 : Study of Eclipses }
\shortauthors{Kudale et al.}
\usepackage{subfig}

\begin{document}

\title{Study of eclipses for Redback pulsar J1227$-$4853}
\author{
Sanjay Kudale\altaffilmark{1},
Jayanta Roy\altaffilmark{1},
Bhaswati Bhattacharyya\altaffilmark{1},
Ben Stappers\altaffilmark{2},
Jayaram Chengalur\altaffilmark{1}}
\altaffiltext{1}{National Centre for Radio Astrophysics, Tata 
Institute of Fundamental Research, Pune 411 007, India}
\altaffiltext{2}{Jodrell Bank Centre for Astrophysics, 
School of Physics and Astronomy, The University of Manchester, 
Manchester M13 9PL, UK}
\affil{}

\begin {abstract}
We present a multi-frequency study of eclipse properties of a transitional 
redback millisecond pulsar J1227$-$4853 discovered in 2014 with the 
GMRT. Emission from this pulsar is eclipsed at 607 MHz for about 37\% 
of its orbit around the superior conjunction. We observe eclipse ingress
and egress transition which last 12\% and 15\% of its orbit respectively, 
resulting in only 36\% of the orbit being unaffected by eclipsing material.
We report an excess dispersion measure (DM) at eclipse boundaries of 
0.079(3) ~pc~cm$^{-3}$ and the corresponding electron
column density (N$_e$) is 24.4(8) x 10$^{16}$ cm$^{-2}$.
Simultaneous timing and imaging studies suggests 
that the eclipses in J1227$-$4853 are not caused by temporal 
smearing due to excess dispersion and scattering but could be caused 
by removal of pulsar flux due to cyclotron absorption of the pulsed 
signal by intra-binary material constraining the companion's magnetic
field. Additionally, near inferior conjunction at orbital phase 0.71
and 0.82 the pulsed emission is significantly delayed which is associated 
with a fading of the pulsed and continuum flux densities. At orbital phase
$\sim$0.82, we measure a change in DM of 0.035(3)~pc~cm$^{-3}$ and  N$_e$ 
of 10.8(8) x 10$^{16}$ cm$^{-2}$ associated with a dimming of up
to $\sim$30\% of the peak flux density.
Such flux fading around a fixed orbital phase is not reported for other
eclipsing binaries. Moreover, this event around inferior conjunction could
be caused by absorption of pulsed
signal by fragmented blobs of plasma generated from mass loss through the
L2 Lagrangian point.
\end {abstract}

\vskip 0.6 cm
\section{Introduction}
\label{sec:intro}
Millisecond pulsars (MSPs) are believed to be generated in a re-cycling 
process where the pulsar accretes mass from its companion star in 
a close binary system resulting in a faster spin period
via transfer of angular momentum (e.g. \cite{Bhattacharya92}).
A special class of fast spinning MSPs (spin period $<$ 8 ms) in 
evolving compact binaries (less than a day), where the pulsar is 
in active interaction with its companion are classified as 
black widow and redback MSP systems. Such compact systems where companions are
ablated away by energetic pulsar winds, are in general referred to as 
spider MSPs. In the majority of such system, the inclination of the
binaries allow the intra-binary 
material to obscure the pulsar emission for a part of its orbit 
resulting in the observed eclipses. The volume occupied by the eclipsing 
material is well outside the companion’s Roche lobe, and 
thus is not gravitationally bound to the companion. 
The energy of an isotropic pulsar wind at the  distance
of the companion is given by $\dot{E}/a^{2}$, where $\dot{E}$
is the spin-down energy of the pulsar and $a$ is the distance to the
companion.
$\dot{E}/a^{2}$ in redback and black widow pulsars are 
$\sim$ 10$^{34}$ $erg/s/R_{\odot}^{2}$, whereas 
$\dot{E}/a^{2}$ for canonical MSPs is around 
$10^{29} - 10^{30}$ $erg/s/R_{\odot}^{2}$. 
\cite{Roy15} reported the discovery of a 1.69 millisecond pulsar
J1227$-$4853, at a dispersion measure (DM) of 43.4 pc~cm$^{-3}$
associated with LMXB XSS J12270$-$4859, using the GMRT at 607 MHz.
PSR J1227$-$4853 is in a 6.9
hours orbit with a companion of mass 0.17$-$0.46 $M_\odot$ and is
eclipsed for large fraction of its orbit at 607 MHz. 
\par
The majority of black widow and redback pulsars exhibit long eclipses
($>$10\% of the orbital period) near their companion's superior 
conjunctions. \cite{Thompson94} gives
a detailed prescription for investigation of the 
eclipse mechanism in such systems.
However, the detailed study of the eclipse properties have
been performed for only a few of the spider pulsars:
PSR J1544$+$4937 \citep{Bhattacharyya13},
PSR B1744-24A (\cite{Lyne90}, \cite{Nice92}, \cite{Bilous19}), 
PSR J1810$+$1744 \citep{Polzin18}, PSR J1816+4510 \citep{Polzin20}, 
B1957$+$20 (\cite{Fruchter88}, \cite{Ryba91},
\cite{Main18}, \cite{Li19}), J2051$-$0827
(\cite{Stappers96}, \cite{Polzin20}. This could be due to 
lack of availability of sensitive instruments operating 
at low frequencies, where the effects of eclipses are expected to be larger.
This is addressed by some of the more recent studies (e.g. \cite{Main18},
\cite{Li19}, \cite{Polzin18} and \cite{Polzin20})  with 
sensitive observations using the Arecibo,
LOFAR, upgraded GMRT (uGMRT) and the Parkes telescope.
\par
In this paper we present a detailed study of the eclipses in the PSR 
J1227$-$4853 system at multiple frequencies. Section \ref{sec:obs-analysis}
details the observations and analysis procedure. Different sub-sections
of Section \ref{lab:results} presents the results from study of the eclipse properties 
of PSR J1227$-$4853. Section \ref{lab:eclipse_591-624} concentrates on main eclipses at 
607 MHz. Investigation of frequency dependent eclipsing is 
presented in Section \ref{sec:dual_freq}.  In addition to the main eclipse, we 
also observe excess dispersion around inferior conjunction, which 
is reported in Section \ref{sec:excess-DM}.
Flux fading observed at eclipse ingress
and around inferior conjunction is reported in Section \ref{sec:cont_pulsed_flux}.
Discussions on these results and a summary are presented in Section \ref{sec:Discussion}. 
\section{Observation and analysis}
\label{sec:obs-analysis}
Following the discovery, PSR J1227$-$4853 is being regularly 
observed using the GMRT coherent array at 607 MHz. 
Most of the observations reported in this paper
were carried out with the legacy GMRT system using GMRT Software Back-end (GSB; 
\cite{Roy10}). We generated filter-bank data products having 512$\times$0.0651 MHz 
channels at 61.44 $\mu$s time resolution.  These data were incoherently 
de-dispersed at the pulsar DM and folded with 
the ephemeris using PRESTO \citep{Ransom02}. We used a multi-Gaussian 
template for extracting times-of-arrival (TOAs) at each observing 
epoch. The TOAs were generated with typically $\sim$4 minutes integration 
time for achieving optimal signal-to-noise (S/N) as well as
a time resolution sufficient 
to probe eclipse transition. Similar time resolution was used in the
imaging  analysis (described below).  PSR J1227$-$4853 is eclipsed for
around 2.8 hours, which is $\sim$ 40\% of 
its orbit \citep{Roy15}.
Many of the timing observations typically of $\sim$1 hour duration 
regularly performed with the GMRT, partially samples eclipse phase
allowing us to probe eclipse characteristics of PSR J1227$-$4853.
\par
In order to probe the frequency dependence of eclipse characteristics 
we observed PSR J1227$-$4853 simultaneously at 300$-$500 and 
550$-$750 MHz using the upgraded GMRT (uGMRT; Gupta et al. 2018).
The increase of instantaneous band-width compensates for the 
reduction of the coherent array gain compared to our earlier
observations resulting from splitting 
antennas into two sub-arrays. The 550$-$750 MHz data was 
recorded in 4096$\times$0.0488 MHz filter-bank output at 
81.92 $\mu$s time resolution, which was incoherently 
de-dispersed and folded. Whereas 300$-$500 MHz data was 
recorded in 512$\times$0.390 MHz coherently de-dispersed 
filter-bank format at 10.24 $\mu$s time resolution in order to 
avoid residual dispersion smearing reducing the TOA uncertainties.
\par
Visibility data were recorded with $\sim$2 
seconds time resolution in parallel with the beam-formed data.  Every 
observation of the target pulsar is accompanied with observation 
of the phase calibrator 1154$-$350 which is sufficiently close 
and strong enough to perform bandpass and gain calibration (7.8 Jy). 
Continuum imaging analysis is carried using an automated imaging 
pipeline (Kudale, Chengalur, Mohan, in preparation) which is 
composed of {\it flagcal} \citep{Chengalur13}, {\it PyBDSM} 
\citep{Mohan15} \& {\it CASA}\footnote{http://www.casa.nrao.edu}.
In total three self calibration and imaging cycles are carried 
out, of which the first two cycles of gain calibration were done with 
phase-only calibration and last was done with amplitude and 
phase calibration.  Final imaging after the last self-calibration 
cycle is done only for the duration for which pulsar was in the 
non-eclipsing phase of its orbit. This enabled us to estimate 
average flux density on given observation epoch.  The self-calibrated 
uvdata were then used to generate snapshot images of the pulsar 
with an average time  duration $\sim$3 minutes to generate the 
lightcurve.  Since the pulsar is a point source, we 
use the peak flux density obtained by fitting a 2D Gaussian to the pulsar 
image to estimate pulsar flux density.  This was done using the {\it imfit} 
task of CASA, with same region box around the pulsar used to do
the fit in all image 
frames.  To obtain errorbars on the flux densities we used the 
task {\it imstat} of {\it CASA} to estimate the rms near the pulsar location.
This we feel, is a conservative, but better estimate of the true
uncertainty than the formal error to the peak of the Gaussian fit.
\section{Results}
\label{lab:results}
\begin{figure}[!ht]
\centering
        \includegraphics[width=5.5in,angle=0]{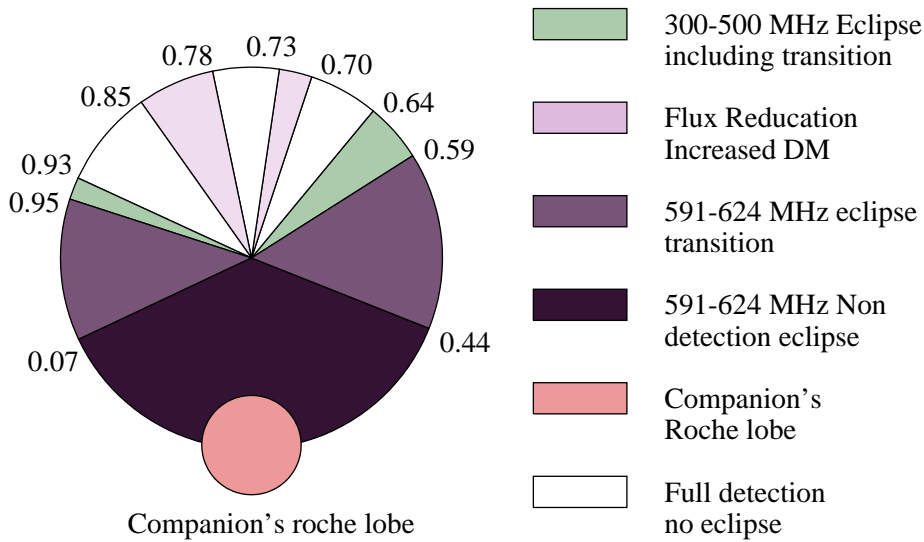}
	\caption[Roche lobe and geometry of Eclipsing binary:top view]
	{Top view of companion's Roche lobe and geometry of eclipsing 
	binary. Companion's Roche lobe and orbit are approximately to 
	the scale assuming the radio timing model \citep{Roy15}.}
	\label{fig:roche_lobe_eclipse}
\end{figure}
Even though the orbital period of PSR J1227$-$4853 is $\sim$6.9 hours, the
source is visible at the GMRT 
sky for only $\sim$3.5 hours. Each observing session, typically of 1 
hour duration, covers only a part of the orbital phase. However, 
regular timing observations performed with the GMRT, allowed us 
to use this collection of observations to probe eclipse boundaries. 
Similar to Fig. 1 by \cite{Polzin19}, we show a schematic 
diagram of the companion's orbit for PSR J1227$-$4853 system highlighting 
companion's Roche lobe (R$_{L}$ = 0.51 R$_{\odot}$, \cite{Eggleton83},
using 1.4 M$_{\odot}$ as pulsar mass and 0.2 M$_{\odot}$ as
companion mass), 
eclipse regions at 591$-$624 MHz and 
300$-$500 MHz. The observed flux fading near inferior conjunction 
associated with increase of the line-of-sight DM is also indicated in
this figure. We describe the main results from multi frequency 
investigation of PSR J1227$-$4853 in the following sections. 
\subsection{Study of eclipses at 591$-$624 MHz}
\label{lab:eclipse_591-624}
\begin{figure}[!ht]
\centering
    \includegraphics[width=7.0in,angle=0]{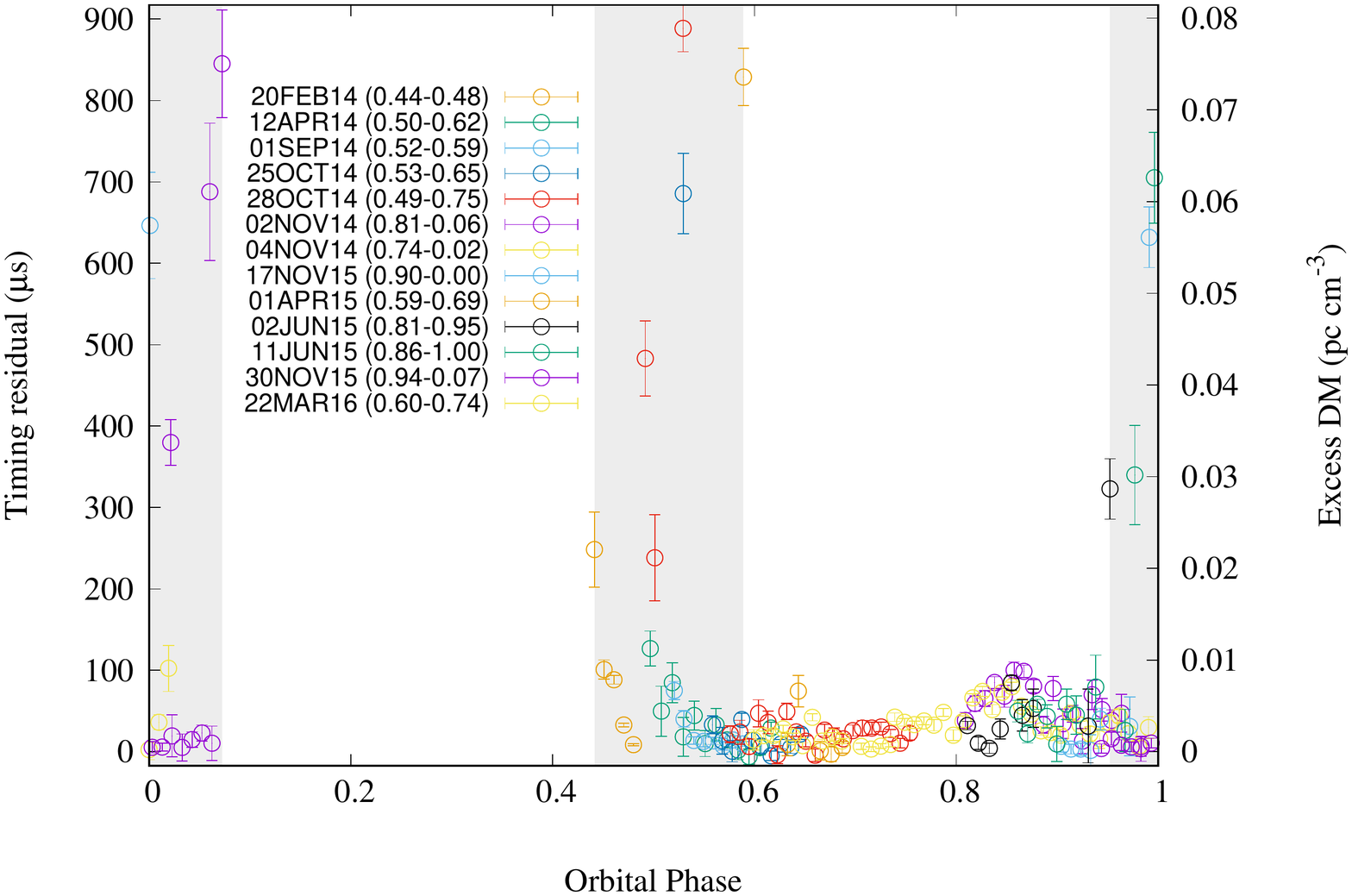}
    	\caption[Variation of timing residuals and excess DM and Ne at 607 MHz.]
	{Variation of timing residual \& DM  with orbital phase 
	on 13 observing epochs (denoted by different colors) at 591$-$624 MHz, 
	individual epochs covering small range of full orbital 
	phase, but collectively  full orbital phase range is covered 
  with all the observations.
	}
	\label{fig:res_DM_Ne}
\end{figure}

Our sample consists of 13 epochs of observations at 591$-$624 MHz, 
out of which 6 observations include an eclipse ingress and 7 
observations include an eclipse egress.  Timing 
residuals of these observations  are presented in Fig. \ref{fig:res_DM_Ne}. 
We observed substantial delays 
in the timing residuals (888(28) $\mu$s) due to 
line-of-sight excess DMs at the eclipse boundaries associated 
with corresponding drops in the flux density. Moreover, we find that
the eclipse ingress and egress transitions are spread over a range of 
orbital phases as shown by shaded regions in Fig. \ref{fig:res_DM_Ne}. 
The eclipse ingress transition starts from $\phi_{B}$ = 0.95 and ends at
0.07, resulting in total ingress  duration of 0.12 in 
orbital phase.  
Using the detection at latest ingress phase ($\phi_{B}$ $\sim$ 0.07)
and earliest egress phase ($\phi_{B}$ $\sim$ 0.44) from a sample of
13 eclipses, we estimate the duration of the completely eclipsed phase
to be 37\% of the orbital period.
 This duration is 
smaller than the value reported in \cite{Roy15}, which was based 
on a single ingress and egress detection. The egress 
transition region is spread over orbital range from 0.44 to 0.59 
resulting in total egress side transition duration of 0.15 in 
orbital phase. Thus the egress transition is seen for longer duration
compared to the ingress transition (by 12.4$\pm$3 minutes ), 
which is also seen in other eclipsing binary systems, e.g. PSR 
J1810$+$1744 \citep{Polzin18}, PSR J1544$+$4937 \citep{Bhattacharyya13}. 
We find the center of the non-detection eclipse (excluding eclipse
transitions) at an orbital phase of 0.255(5), which matches with the superior conjunction orbital phase. The estimated line-of-sight excess DM and 
electron column density (N$_e$) from timing residuals are shown 
in  Fig. \ref{fig:res_DM_Ne}. The full eclipse  and eclipse transition
zones (shaded regions)  seen in Fig. \ref{fig:res_DM_Ne} can also 
be visualized in the schematic top view of the 
eclipse geometry in Fig. \ref{fig:roche_lobe_eclipse}, 
where these regions are highlighted in dark purple and purple colors respectively. 
The maximum delay in timing residuals around eclipse transitions detected 
for PSR J1227$-$4853 is 888(28) $\mu$s at 591$-$624 MHz. This gives 
excess DM of 0.079(3) ~pc~cm$^{-3}$ and N$_e$ of 24.4(8) x 10$^{16}$ cm$^{-2}$ 
(see  Fig. \ref{fig:res_DM_Ne}).
We estimate the corresponding electron density in the eclipse
region (n$_e$ $\sim$ N$_e$/$a$ ) as 1.5 x $10^6$ cm$^{-3}$,
which is at least an order of magnitude higher than the
electron density expected in the stellar wind (according to \cite{Johnstone15}
n$_e$ due to the stellar wind at a distance similar to $a$ is $\sim$  10$^5$ cm$^{-3}$).
This indicates that 
ablation from the companion is significantly contributing to the 
intra-binary material causing
eclipses. 
%
%
This system also exhibits eclipses for a longer fraction of orbital phase. 
We compare the eclipse properties of PSR J1227$-$4853 with the known 
eclipsing binaries in Section \ref{sec:Discussion}.
\subsection {Simultaneous dual frequency study of eclipses}
\label{sec:dual_freq}
\begin{figure}[!ht]
\centering
    \includegraphics[width=6.6in,angle=0]{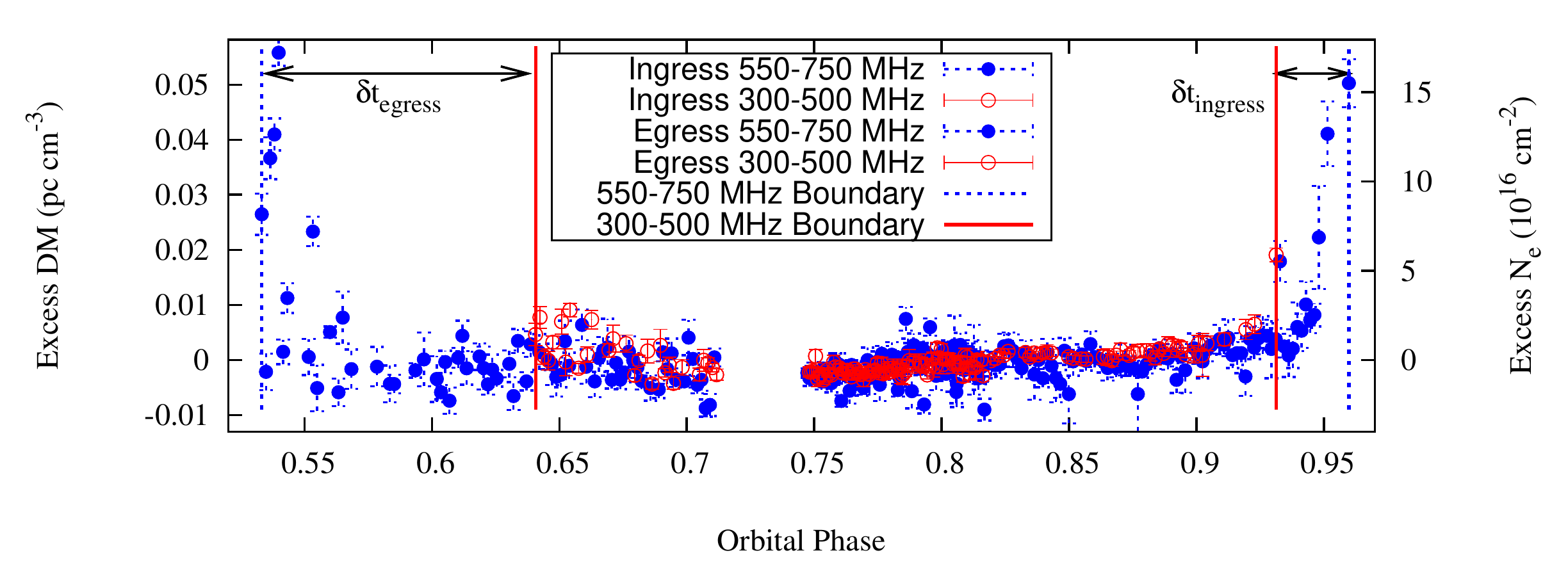}
	\caption[Excess DM \& N$_{e}$ at eclipse egress \& ingress boundaries
	measured at 400 \& 650 MHz]
	{Variation of excess DM \& Ne at eclipse egress \& ingress boundaries 
	measured simultaneously at 300$-$500 MHz (hollow red circles)
	\& 550$-$750 MHz 
	(filled blue circles). One epoch covers egress (24th May 2019) from 
	orbital phase 0.53 to 0.82 and another 
	covers ingress (14th May 2019) from orbital phase 0.80 to 0.96. 
	A break seen in egress observation at orbital 
	phase $\sim$0.72 is due to re-phasing of the array.
	}
	\label{fig:simultaneous_dual_freq_eclipse}
\end{figure}
In order to probe the frequency dependence of the eclipse duration for PSR 
J1227$-$4853 we carried out simultaneous dual-frequency observations at 
300$-$500 and 550$-$750 MHz using the uGMRT. Observations performed on 
14 May 2019 and 24 May 2019 allowed us to probe the eclipse ingress and 
egress transitions respectively. This eliminates the effect of temporal 
variations of eclipse boundaries (as seen in Fig. \ref{fig:res_DM_Ne}) 
while estimating the frequency dependence of the eclipse duration. The 
eclipse region for 300$-$500 MHz is shown by light green color in 
Fig. \ref{fig:roche_lobe_eclipse}. Variation of the  excess DM and N$_e$ 
with orbital phase derived from the best-fit timing residuals are shown in 
Fig. \ref{fig:simultaneous_dual_freq_eclipse}. Eclipse boundaries are 
marked by vertical lines: red for 300$-$500 MHz and blue for 550$-$750 MHz.
We observe a larger eclipse duration at the lower frequency band (i.e.
$\sim$1.2 times longer for 300$-$500 MHz band than 550$-$750 MHz band) and 
we note a possible asymmetry in frequency dependence of eclipse transitions 
in ingress and egress phase in the 300$-$500 MHz band compared to that in 
the 550$-$750 MHz band. The ingress starts earlier, 
$\delta t_{ingress} = $ 11.86$\pm$0.5 minutes, and egress ends later, 
$\delta t_{egress} = $ 44.57$\pm$0.5 minutes, at 300$-$500 MHz.
If we consider a power-law dependence of eclipse duration with frequency 
($T_{eclipse}~  \alpha~  \nu^{n}$), we estimate a power-law 
index of $n$ = $-$0.44 from these simultaneous observations.
Frequency dependent eclipse durations are observed for some of 
the other eclipsing binaries as well.  Earlier studies report 
that at lower frequencies the eclipse duration is seen to be
larger compared  to that of higher frequencies for a given system. 
We have listed excess DM, pulsar wind flux ($\dot{E}/a^{2}$), eclipse
duration and power-law index for eclipsing binaries in Table \ref{tab:literature_ref}.
We observed 
an asymmetry in eclipse boundaries between the two observing bands where
$\delta t_{egress} > \delta t_{ingress}$ by 32.7$\pm$0.7 minutes. 
From this we can derive separate power-law frequency dependence 
for ingress ($T_{ingress}~  \alpha~  \nu^{n_{i}}$) and egress
($T_{egress}~  \alpha~  \nu^{n_{e}}$) transitions (w.r.t superior conjunction), 
where  $n_{i}$ =  $-$0.19 and  $n_{e}$ = $-$0.66.
\begin{figure}
\centering
\begin{tabular}{c c}
\subfloat{\includegraphics[scale=0.4,angle=0]{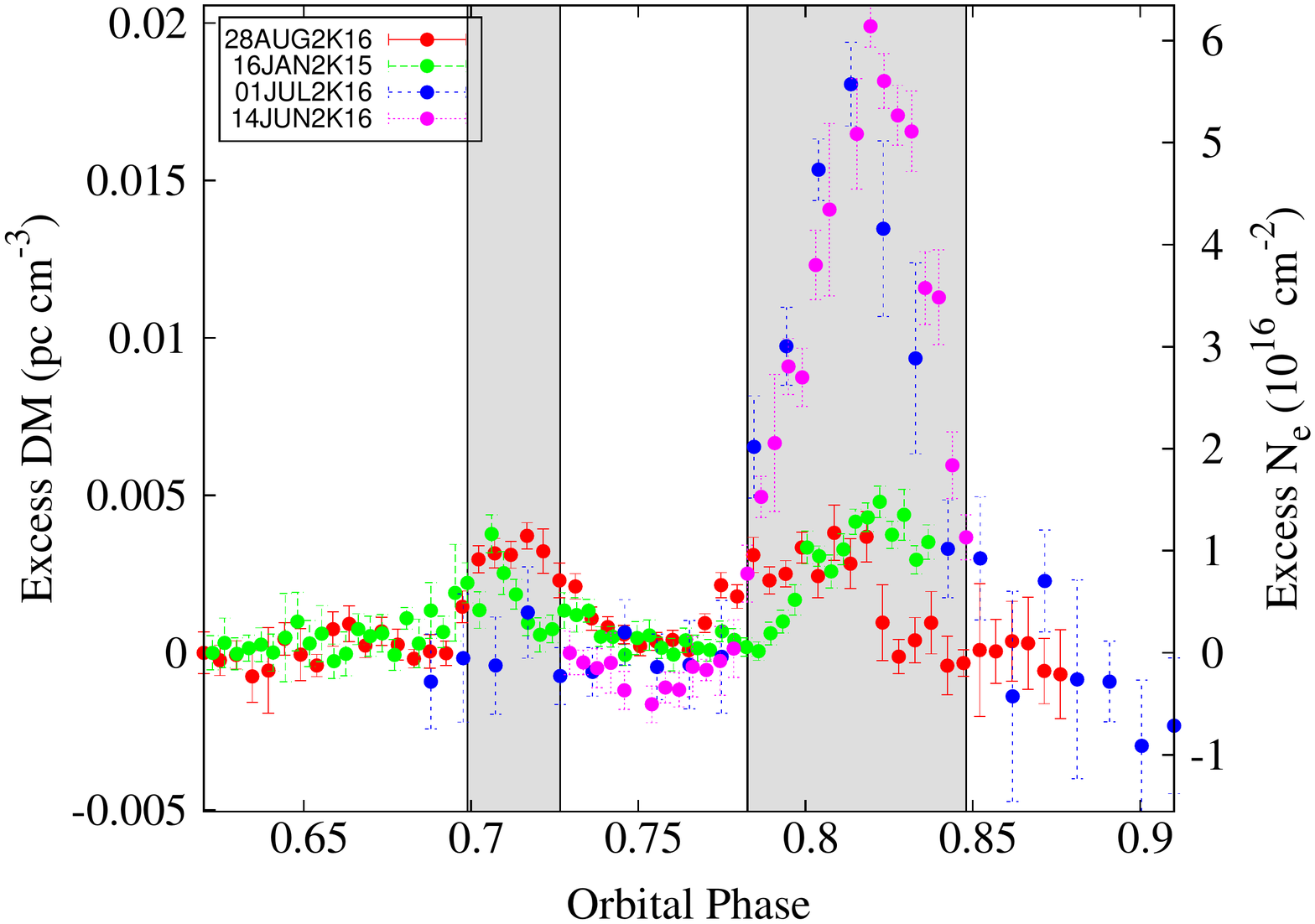}}
\\  
\multicolumn{2}{c}
\subfloat{\includegraphics[scale=0.4,angle=0]{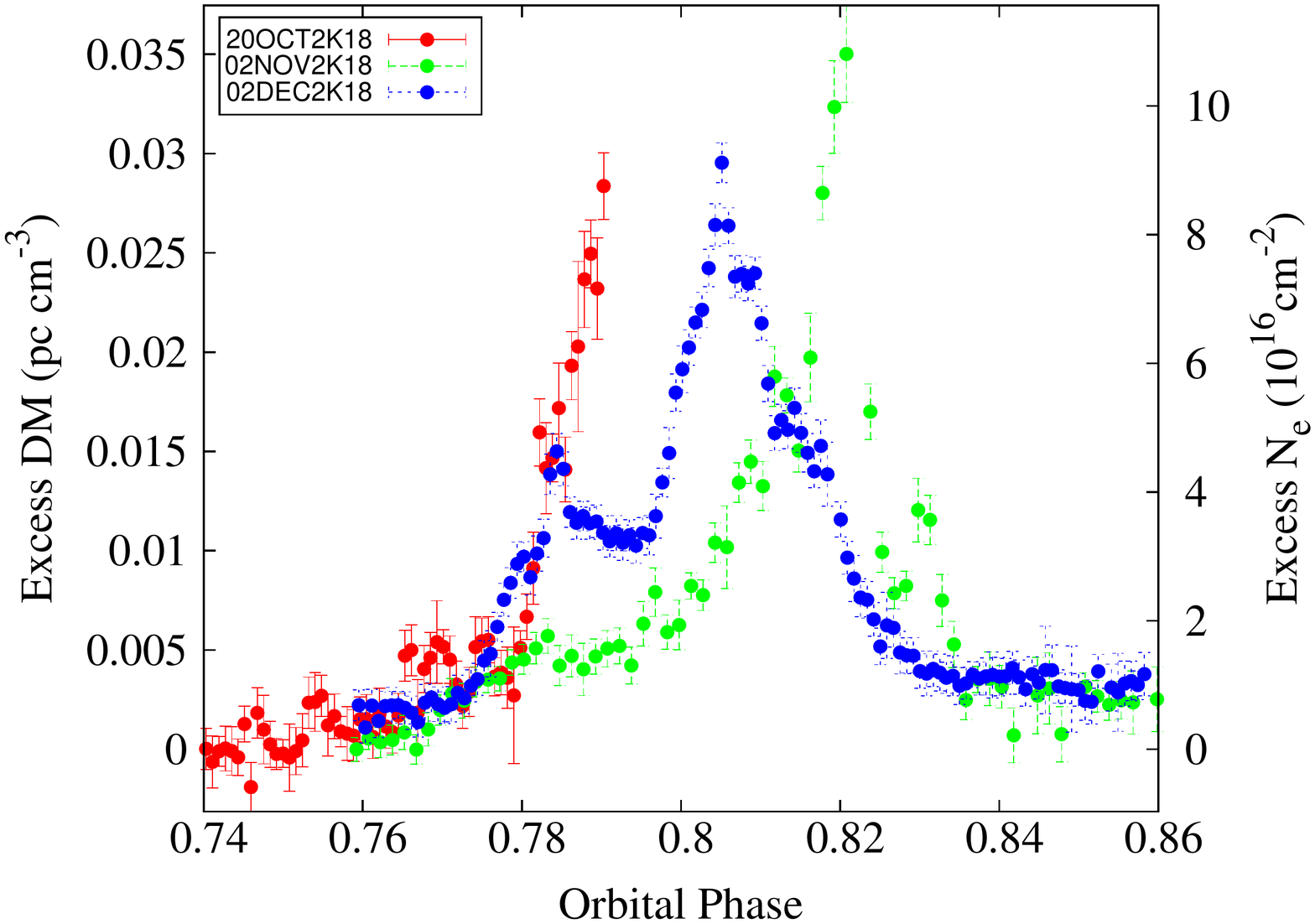}}
\end{tabular}
  \caption[DM and N$_e$ measured at $\phi_{B}$ = 0.7 \& 0.8 at 607 MHz
  with 32 MHz bandwidth \& $\phi_{B}$ = 0.7 at 650 MHz with 200 MHz bandwidth.]
  { Top panel: Variation of excess DM and N$_e$ measured around
  orbital phase 0.7 \& 0.8 at 591$-$624 MHz with time resolution
  of 2.3 minutes. Bottom panel: Variation of excess DM and N$_e$
  measured around orbital phase 0.8 at 550$-$750 MHz with coherently
  de-dispersed observations with time
  resolution of 26 seconds. More sensitive data with higher
  time resolution brings out the pattern of variation of excess N$_e$ with orbital phase.
  }
  \label{fig:res_DM_Ne_SE}
\end{figure}
\subsection {Excess dispersion around inferior conjunction} 
\label {sec:excess-DM}
In addition to eclipses seen at orbital phases from 0.95 up to 0.59, 
PSR J1227$-$4853 exhibits occasional occurrences of residual delays 
around $\phi_{B}$ of $\sim$0.7 and $\sim$0.8 
(marked by the light purple color in the eclipse geometry in
Fig. \ref{fig:roche_lobe_eclipse}). This is well outside the 
eclipse regions, centered around inferior conjunction, $\phi_{B}$ = 0.75 
(seen in the top panel of Fig. \ref{fig:res_DM_Ne_SE}).
The largest excess DM and N$_e$ we measured at $\phi_{B}$ = 0.82 
is 0.0199(6) ~pc~cm$^{-3}$ and 6.1(2) x 10$^{16}$ cm$^{-2}$ 
respectively, which is factor of 4 lower than the values 
measured at the eclipse boundaries. Whereas at $\phi_{B}$ = 0.71 
we measured an excess DM of 0.0037(6) ~pc~cm$^{-3}$ and N$_e$ 
of 1.1(2) x~10$^{16}$ cm$^{-2}$. The durations of these 
phenomenon of excess dispersion measured at $\phi_{B}$ 
of 0.7 and 0.8 are 11.5$\pm$1.7 and 27.1$\pm$1.7 minutes respectively. 
We present three epochs of coherently de-dispersed observations at 550$-$750 
MHz probing excess dispersion around $\phi_{B}$ of 0.82 at higher 
time-resolution as seen in the bottom panel of Fig. \ref{fig:res_DM_Ne_SE}. 
The higher S/N data from the coherently de-dispersed observations 
allow us to probe eclipses at time resolution of 26 seconds as compared 
to the 591$-$624 MHz incoherently de-dispersed  data which has time 
resolution of 2.3 minutes. These observations equipped with 
higher time-resolution and enhanced sensitivity (due to wider 
band-width) reveal a maximum excess DM of 0.035(3) pc~cm$^{-3}$ 
and N$_e$ of 10.8(8) x 10$^{16}$ cm$^{-2}$.
We estimate the corresponding electron density in the eclipse
region (n$_e$ $\sim$ N$_e$/$2a$ ) as 0.3 x $10^6$ cm$^{-3}$,
which is at least an order of magnitude higher than the
electron density in the stellar wind (according to \cite{Johnstone15}
n$_e$ due to stellar wind at a distance similar to separation between 
companion and inferior conjunction is $\sim$ 10$^4$ cm$^{-3}$).
\subsection {Continuum and pulsed flux} 
\label{sec:cont_pulsed_flux}
\begin{figure}
\centering
\begin{tabular}{c c}
\subfloat{\includegraphics[scale=0.4,angle=0]{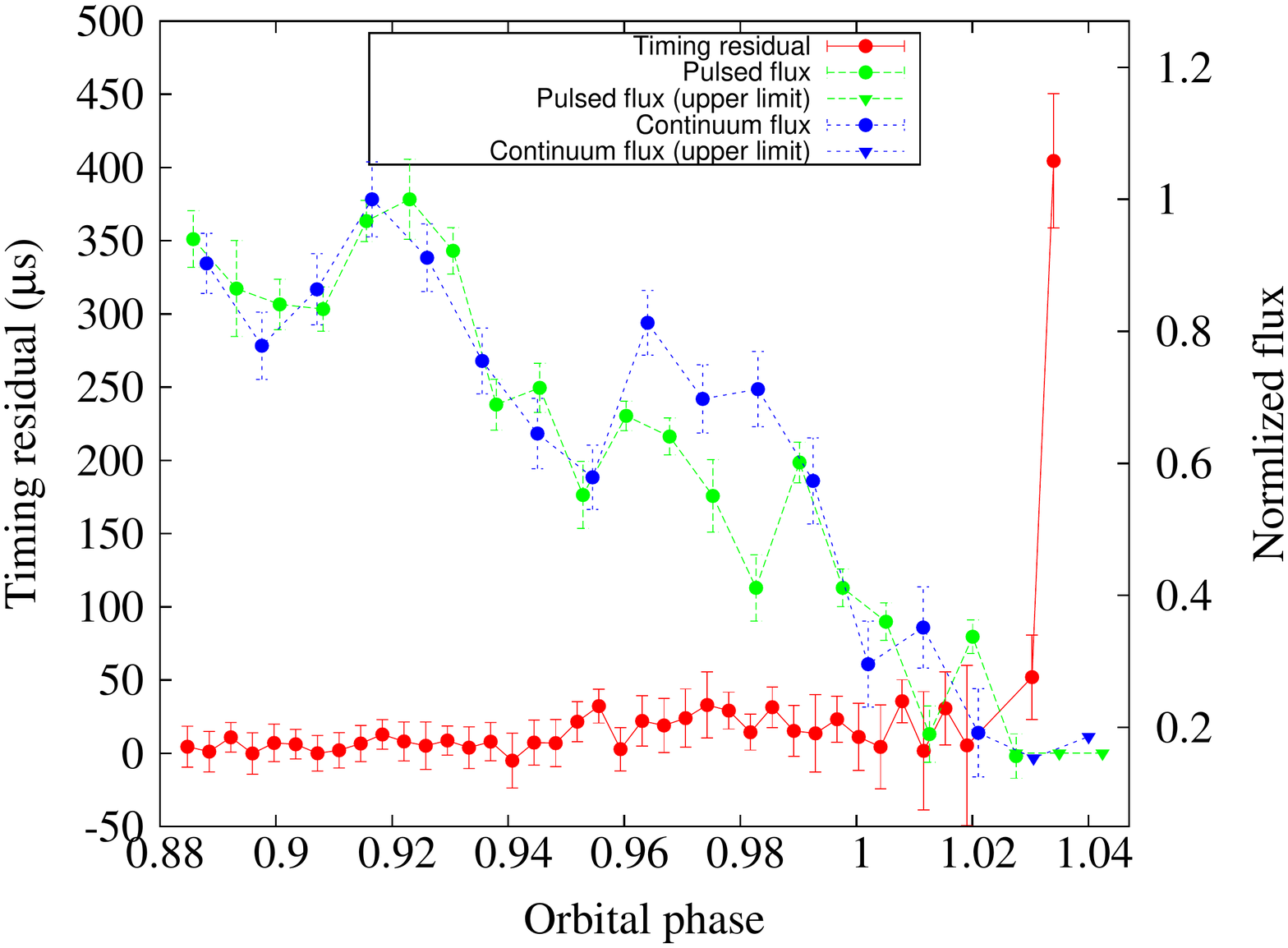}}
\\
\multicolumn{2}{c}
\subfloat{\includegraphics[scale=0.4,angle=0]{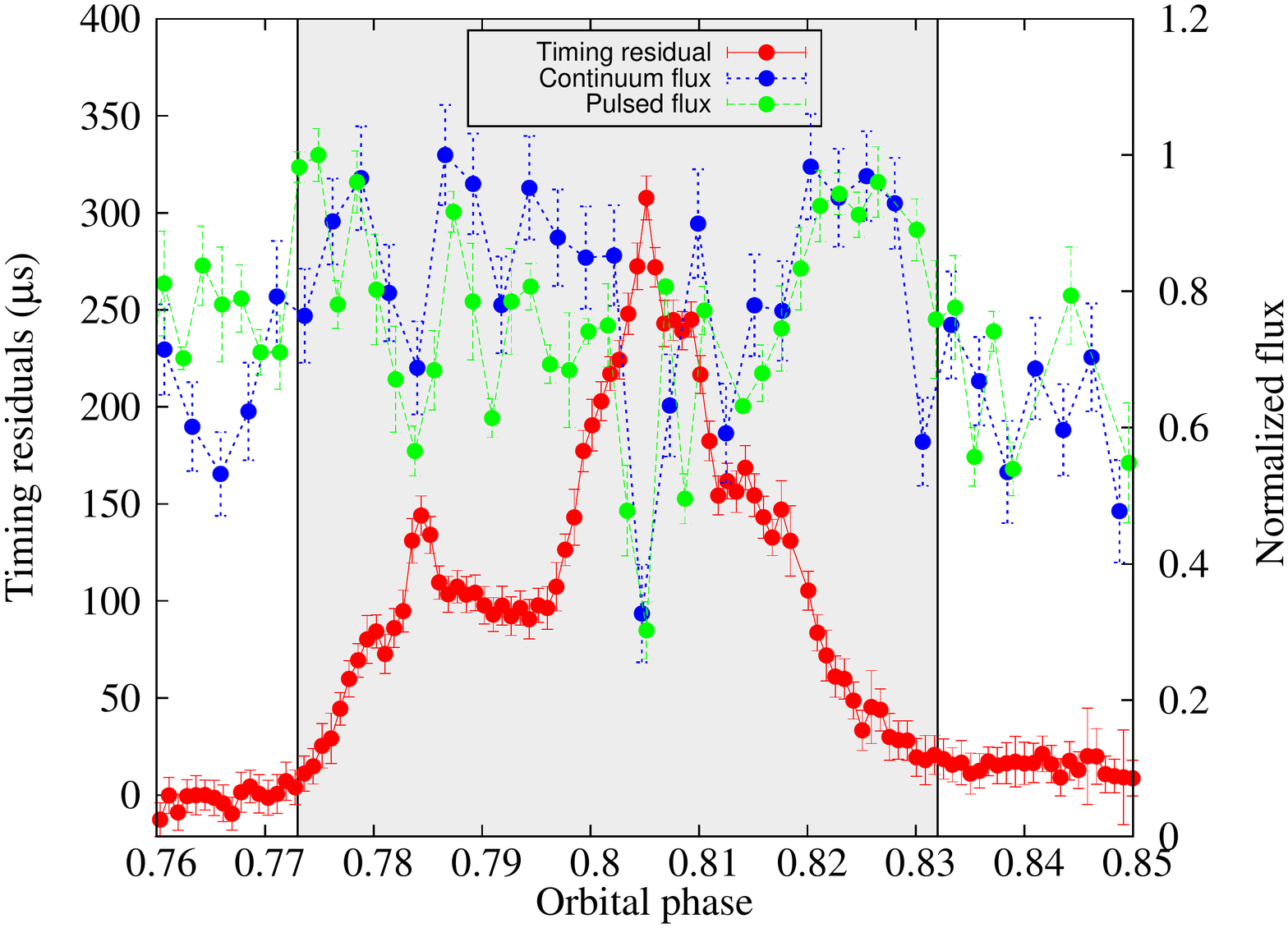}
}
\end{tabular}
  \caption[Timing residuals, Beam and Imaging flux around inferior
  conjunction and during ingress]
  {Top panel: Variation of timing residuals (added to mark start of eclipse)
  and flux densities (both pulsed and continuum)  with orbital phase during 
    ingress at 550$-$750 MHz observed on 01 Jan 2019.
  Bottom panel: Variation of timing residuals and continuum flux densities
  with orbital phase around inferior conjunction at 550$-$750 MHz
  observed on 02 Dec 2018. The highlighted region
  indicates the anti-correlated variation of continuum flux with timing
  residuals.}
  \label{fig:Beam_image_flux}
\end{figure}
Aided by the capability of simultaneously recording visibilities along 
with the tied-array coherent beam from the 
GMRT interferometer, we estimated 
flux densities around eclipse boundaries using continuum imaging 
and compared that with the pulsed flux densities. Unlike pulsed flux 
densities the flux densities obtained from continuum imaging are expected 
to be unaffected by the temporal smearing caused by excess dispersion 
and/or scattering. This comparative study of continuum  and pulsed 
flux densities can be used for understanding the eclipse mechanism, 
which was done by \citep{Roy15} for PSR J1227$-$4853 while probing the egress 
boundary. Apart from PSR J1227$-$4853, imaging studies for eclipsing 
binaries were done by \cite{Polzin20} for PSRs B1957$+$20 \& J1816$+$4510, 
by \cite{Polzin18} for PSR J1810$+$1744 
and by \cite{Broderick16} for PSR J2215$+$5135.
\par
For PSR J1227$-$4853 we analyzed the imaging data (details 
in Section \ref{sec:obs-analysis}) to produce a
lightcurve for the pulsar signal during eclipse ingress and 
during the instances of excess dispersion  around inferior 
conjunction. The top panel of Fig. \ref{fig:Beam_image_flux} shows
the continuum (marked in blue) and pulsed   (marked 
in green) flux densities as function of orbital phase for 
an eclipse ingress. A curve showing the timing residuals (marked in 
red) reaching up to 404 $\pm$ 46 $\mu$s is also added in 
this plot. The continuum and pulsed flux densities show
correlated changes as the pulsar is transitioning 
into eclipse at $\phi_{B}$ $\sim$0.03, where the timing 
residuals are rapidly increasing. We have carried out 
similar lightcurve analysis for the 2 Dec 2018 event of 
excess dispersion around $\phi_{B}$ = 0.8. As seen 
in the bottom panel of Fig. \ref{fig:Beam_image_flux}, 
the TOAs (red curve) are delayed by about
307 $\pm$ 11 $\mu$s at $\phi_{B}$ = 0.8. The continuum 
flux density (blue curve) and pulsed flux density (green curve) fade in  anti-correlation 
with the arrival times. The region with the excess 
dispersion delay is highlighted in bottom panel of 
Fig. \ref{fig:Beam_image_flux}. The two peaks of timing 
residual at $\phi_{B}$ = 0.783 and 0.805 are  exactly
coinciding with two dips of the continuum and pulsed flux densities.
The observed continuum flux density at $\phi_{B}$ = 0.805
is around 30\% of the peak continuum flux density measured 
at orbital phase $\phi_{B}$ =0.825.
Overall the pulsed flux density is consistent with the continuum flux density. 
For the orbital range of 0.79 to 0.8 the pulsed flux density is a little bit lower
than the continuum flux density. This could be due to the presence of temporal 
broadening caused by increased DM as also seen in \cite{Polzin20}. 
We performed light curve 
measurements of nearby point sources within the field-of-view 
showing no significant variation of flux densities over the
observing span.
We observed scintles of a few MHz in size, as well as flux brightening
of a few minutes duration on both sides of the event of excess dispersion 
(orbital phase $\sim$0.805), which can explain the enhancement of flux
densities (highlighted region in the bottom panel of Fig. \ref{fig:Beam_image_flux}).
%
\section{Discussion and Summary}
\label{sec:Discussion}
\begin{table}[h!]
\caption{Parameters for eclipsing binary millisecond pulsar systems 
listed in column 1
indicating its type, redback (RB) or black widow (BW).
 Column 2 presents the excess dispersion.
 Column 3 presents $\dot{E}/a^{2}$ where $\dot{E}$ is spin-down energy of the pulsar 
 and $a$ is distance to the companion.
 Column 4 presents eclipse duration with corresponding  frequency in parenthesis. 
 Column 5 denotes the index of power-law dependence (n) of full eclipse duration with frequency.
}
\begin{tabular}{l l l l l l}
\hline
Pulsar & Excess DM    & $\dot{E}/a^{2}$$^{\alpha}$ (10$^{35})$  & Eclipse  & n & Reference$^\zeta$ \\
Name   & (pc~cm$^{-3})$ & ($erg/s/R_{\odot}^{2})$ & duration$^{\beta}$ &   &\\
\hline
J1023$+$0038 (RB) & 0.15(700) &0.33 &40(685) & -0.41 & 1\\
J1048+2339 (RB) & 0.008(327) & 0.03 & 57(327) & $-$ & 10\\
J1227$-$4853(RB) & 0.079(607) & 0.29 & 64(607) &  -0.44     & 2\\
J1227$-$4853$^{\gamma}$ (RB) & 0.035(607) & $-$ & 6(607) &   $-$ & 2\\
J1544$+$4937 (BW) & 0.027(607) & 0.11 & 13(322) & $-$  & 3\\
J1723$-$2837 (RB)  & $-$ & 0.04 & 26(1520) & $-$ & 4\\
B1744$-$24A (RB) & 0.6(1499.2) & $-$ & $\sim$50$^\delta$(820) & $-$ & 5\\
J1810$+$1744 (BW) & 0.015(325) & 0.18 & 13(149) & -0.41 & 6\\
J1816$+$4510 (RB) & 0.01(149) & 0.08 & 24(121) & -0.49$^\epsilon$ & 8 \\
B1957$+$20 (BW) & 0.01(149) & 0.22 & 18(121) & -0.18 & 8\\
J2051$-$0827 (BW)  & 0.13(705-4023) & 0.06 & 28(149) & -0.41  & 7,8\\
J2215$+$5135 (RB) & $-$ & 0.28 & 66(149) & -0.21$^\epsilon$ & 8,9\\
\hline
\end{tabular}
\\ \\
${\alpha}$ : Using  https://apatruno.wordpress.com/about/millisecond-pulsar-catalogue/\\
${\beta}$ : The eclipse duration (in \% of orbit) includes non-detection and
associated ingress, egress transition.\\
$\gamma$ : Parameters for excess dispersion observed around inferior conjunction.\\
$\delta$ : For majority of the observed eclipses. However, 
  observed eclipse durations are variable and sometimes completely 
  enshrouding the pulsar \citep{Bilous19}.\\
$\epsilon$ : The value of the estimated power law index using all 
  available frequency measurements as given in the recent literature.\\
$\zeta$ : List of references; 1: \cite{Archibald09}; 2: Current work; 3: \cite{Bhattacharyya13}; 4: \cite{Crawford13}; 5: \cite{Bilous19}; 6: \cite{Polzin18}; 7: \cite{Polzin19}; 8: \cite{Polzin20}; 9: \cite{Broderick16}, 10: \cite{Deneva16}\\
 \label{tab:literature_ref}
\end{table}
\par
We find that during ingress and egress the pulses are significantly 
delayed relative to best-fit timing model. The largest timing residual 
deviation that we measure is 888(28) $\mu$s. We 
estimate excess DM and N$_e$ as 0.079(3) ~pc~cm$^{-3}$ and 
24.4(8) x 10$^{16}$ cm$^{-2}$ respectively.
The eclipse duration including ingress, egress transition for PSR J1227$-$4853 is about 265$\pm$3 minutes (64\% of its orbit), indicating that for a larger fraction of its orbit the pulsar is enshrouded by the intra-binary materials.
From Table \ref{tab:literature_ref} we note that the observed values of excess DM, N$_e$, eclipse duration and $\dot{E}/a^2$ 
for PSR J1227$-$4853 are similar to PSR J1023$+$0038, which is the other LMXB$-$MSP transitioning system.
An asymmetry is seen between egress and ingress duration, 
egress being longer by 12.4$\pm$3 minutes. This asymmetry can be caused by a tail of eclipsing 
material swept back due to orbital motion of companion, which is
also observed for other eclipsing binaries (e.g.  \cite{Polzin20}).
Such asymmetries 
can be generated by the interaction of out-flowing gas 
from companion with pulsar radiation, which can create 
increased density in the trailing part of the outflow 
as shown using hydrodynamical simulation by \cite{Tavani91}
suggesting this as the explanation for 
the observed eclipses of PSR B1957$+$20 \citep{Fruchter90}.
From the dual frequency observations on two epochs one covering egress 
boundary and another covering ingress boundary, we observed that 
300$-$500 MHz eclipse duration is longer than 550$-$750 MHz. A longer
eclipse duration at lower frequencies is also observed for other
eclipsing binaries (\cite{Broderick16}, \cite{Polzin18},
\cite{Stappers01}, \cite{Polzin20}). In addition we observe 
that for PSR J1227$-$4853 ingress
boundary starts earlier ($\sim$11.9 minutes) and egress 
ends later ($\sim$ 44.6 minutes) at lower frequency (300$-$500 MHz)
than at higher frequency (550$-$750 MHz), i.e
$\delta t_{egress} > \delta t_{ingress}$ by 32.7$\pm$0.7 minutes. 

%
%
We estimate the power law index for the frequency dependent eclipse
duration as $n$ = $-$0.44.
From Table \ref{tab:literature_ref}, we find generally redback pulsars have 
relatively longer eclipse duration and excess DM at the 
eclipse boundaries compared to the black widow systems.
Future study of a statistically significant sample of such 
eclipsing binaries over a wide frequency range is warranted for better understanding.
\par
We observe a  fading of the pulsar flux density 
around inferior conjunction ($\phi_{B}$ $\sim$ 0.7 \& 0.8) 
which is also associated with an
excess timing delay on several occasions ($\sim$ 25\% of all observations).
To our knowledge such systematic change of flux
density  around a fixed orbital phase  (i.e. inferior conjunction in this case)
is not reported for any other eclipsing binary.
Occasional clustering of fragmented blobs of 
plasma around the  inferior conjunction could possibly lead 
to such decrease in flux. 
The maximum value of excess DM and N$_e$ observed
around inferior conjunction
($\phi_{B}$ = 0.82) is factor of two to four times lower than that 
observed at the eclipse boundary for PSR J1227$-$4853.
In this context we note that for PSR J1544$+$4937 
(having very similar excess DM and N$_e$ as seen in PSR J1227$-$4853)
frequency dependent eclipsing around superior conjunction is observed,
where the pulsed signal exhibits a decrease in flux at higher 
frequency ($\sim$ 607 MHz) and is completely eclipsed at 
lower frequency ($\sim$ 322 MHz) as reported  by \cite{Bhattacharyya13}.
Future investigations of PSR J1227$-$4853
at lower frequencies may reveal frequency dependent eclipsing 
around inferior conjunction.
Short eclipses are generally seen around eclipse region centered on
superior conjunction in other eclipsing binaries, e.g. 
for PSR J1544$+$4937 by \cite{Bhattacharyya13}. However for 
PSR J1227$-$4853 we observe the phenomenon of excess dispersion with flux
fading preferentially centered around inferior conjunction.
According to \cite{deMartino15}, the X-ray emission originates in an intra-binary shock produced by the interaction of the 
outflow from the companion and the pulsar wind.
We also note that \cite{deMartino15} observed a dip in the count rate centered at
$\phi_B$ = 0.75 while monitoring X-ray orbital modulation 
of the pulsar. Radio observations 
reported in this paper have at least an order of
magnitude better orbital phase  resolution than  the X-ray observations, 
which possibly allowed us 
to resolve the single dip seen in X-ray in two symmetric dips observed 
in radio around inferior conjunction.
\cite{Tavani91} explained the observed eclipse properties 
for PSRs B1957$+$20 and B1744$-$24A using hydrodynamical simulations 
of the companion's wind outflow. They showed that the eclipses are created 
due to the shocks generated by interaction between the pulsar radiation and 
the out flowing gas from the companion star. They explained drastic eclipse 
changes observed for PSR B1744$-$24A by \cite{Lyne90}, while inferring 
that the eclipse shape is dependent on the thermal and kinetic state of the 
out flowing gas which could be time variable. By progressively decreasing
mass loss rate \cite{Tavani93} arrived at a final mass configuration
allowing the pulsar to be visible for a large fraction of orbital phase. Whereas 
for progressively increasing or for a constant but relatively large value of the 
mass loss rate, pulsar could get completely enshrouded. 
 According to \cite{Linial17} mass transfer through L2 Lagrangian 
point could happen for a system having rapid orbital evolution. In case of 
PSR J1227$-$4852 mass transfer during accretion phase through L2 could
be responsible for material floating around inferior conjunction causing
excess dispersion.
The observed occasional 
flux fading around the inferior conjunction for J1227$-$4853 could also be
caused by systematic changes in final mass configurations achieved via
variations in the mass loss rate or other parameters such as temperature or
Mach number. Frequent multi-frequency observations are planned to probe this
in more detail.
\par
From simultaneous timing and imaging analysis we find pulsed and continuum 
flux densities of PSR J1227$-$4853 follow a similar trend at eclipse
ingress. \cite{Roy15} reported similar finding at eclipse egress for the same
pulsar.
In earlier studies the decrease of continuum flux densities at
eclipse boundaries were reported by \cite{Polzin18} for 
PSR J1810$+$1744 and by \cite{Broderick16} for PSR J2215$+$5135.
We also measure the variations of continuum flux densities around
inferior conjunction (presented in Section \ref{sec:excess-DM}) 
and find that minima in continuum flux density coincides 
with the maxima in excess dispersion. 
\par
Now we investigate possible eclipse mechanisms following \cite{Thompson94}.
In order to study the pulse smearing due to dispersion as a cause of the eclipse, DM~$\sim$1.3~pc~cm$^{-3}$~is required to disperse pulsed emission completely. However, the
measured largest excess DM at eclipse 
boundary is  0.079(3)~pc~cm$^{-3}$ which 
is a factor of $\sim$16 less and hence 
rule out the dispersion as the 
cause of eclipse. The scattering due to excess N$_e$ can broaden the pulse
and change pulse profile. However, we have not seen any signature of profile
evolution at eclipsing boundaries. Thus scattering as a cause is ruled out.
Moreover, temporal smearing due to the dispersion or scattering 
is not expected to change the  continuum flux density.
For refraction to be the cause of the eclipse the expected
group delay at the ingress or egress would be $\sim$10$-$100 ms
as reported by \cite{Thompson94}.
We measure maximum time delay around eclipse boundary
$\sim$888 $\mu$s, for PSR J1227$-$4853,  
which is at least an order of magnitude 
smaller than the  group delay required for refraction of radio beam 
causing eclipse. This implies refraction can not be the
cause of eclipse.
 For an eclipsing binary system with temperature T and clumping factor of the eclipsing medium
 $f_{cl}$ ($f_{cl} = <n_{e}^{2}>/<n_e>^{2}$), the optical depth due to
 free-free absorption is given by Equation \ref{eq:tau_ff}  \citep{Thompson94}, 
 where $N_e$ is electron column density, L is absorption length.

 \begin{equation}
\label{eq:tau_ff}
\tau_{ff} \simeq 3.1 * 10^{-8} { {f_{cl}} \over ~ {T_{7}^{3/2}}} { ~ N_{e,17}^2} { ~ L_{11}^{-1}}
\end{equation}
Using Equation \ref{eq:tau_ff}
we derive, T $\leq$ 10$^2$ x  $f^{2/3}_{cl}$ K, as relation between the temperature T and clumping factor of the eclipsing medium.  
This indicates that for free-free absorption to be the cause of eclipse (i.e. $\tau_{ff} > 1$) 
in PSR J1227$-$4853 with N$_e$ = 24.4 x 10$^{16}$ cm$^{-2}$   at 
eclipse boundary and absorption 
length about twice the size of the eclipse zone, we need either very
high clumping factor or very low temperature.
%
Assuming a temperature range from an unheated star to an irradiated star 
(i.e. 5500 $-$ 500000 K) according to \cite{deMartino14}, we expect the
range of clumping factor to be 400 to 3.5 x 10$^5$, which is not physically
possible \citep{Muijres12}, ruling 
out free-free absorption as the cause of eclipse.
%
Considering PSR J1227$-$4853 has an average
flux density at 650 MHz (S$_{\nu}^{0}$) $\sim$1.2 mJy, spectral index ($\alpha$) $\geq$ $-$1.8 and distance (d$_{kpc}$) $\sim$1.4 kpc, demagnification (M) $\sim$ $(R_c/2r)^2$, where $R_c$ is radius of curvature
of plasma cloud and r is distance from center  of curvature, the induced Compton optical depth can be calculated with Equation \ref{eq:tau_ind} \citep{Thompson94}. 

\begin{equation}
\label{eq:tau_ind}
\tau_{ind} \simeq {4 * 10^{-5}} ~ {{ N_{e,17}^2 ~ S_{\nu}^{0}} \over {\nu_{9}^{2} }} ~ |\alpha + 1 |~ \Bigg({{d_{kpc}} \over {a_{11}}}\Bigg)^{2} M
\end{equation}

We calculate 
the upper limit of induced Compton depth $\tau_{ind}$ $\leq$ 7.6 x $10^{-5}$, which rules out induced Compton scattering to be the cause of the eclipse.
The decrease of continuum flux density at eclipse boundary as well as
flux fading around inferior conjunction indicates
absorption of pulsar flux by line-of-sight material could be a plausible
cause of eclipse.
In order to check if cyclotron-synchrotron absorption of pulsar
emission by non-relativistic or relativistic electrons is the cause of the
eclipse we estimate the magnetic field of the  eclipsing plasma with Equation \ref{eq:mag_field} \citep{Thompson94} where
m = $\nu$/$\nu _{B}$, $\omega_{B} = 2\pi \nu_{B} = eB/m_ec$.
\begin{equation}
\label{eq:mag_field}
B = 350 m^{-1} \nu_{9} ~~ G
\end{equation}
We calculate the 
magnetic field in the vicinity of the companion to be
27G, and the cyclotron fundamental frequency to be 77 MHz.
Observed eclipses reported in this paper for PSR J1227$-$4853 are centered at 
322 \& 607 MHz which are 
4th and 8th harmonics of this cyclotron fundamental frequency. 
In this context, we note that  eclipses for
PSR J1544$+$4937 have been seen up to 20th 
harmonic of its fundamental cyclotron frequency \citep{Bhattacharyya13}.
For PSR J1227$-$4853 cyclotron absorption at fundamental cyclotron frequency
and its lower harmonics can be the cause of eclipse. 
The observed larger frequency dependence of the eclipse egress
compared to the ingress can also be  explained by the the presence 
of more stellar material around the eclipse egress than ingress which 
could result into enhanced frequency dependence of cyclotron absorption optical depth. 
Since cyclotron absorption
optical depth decreases for higher harmonics, it will
be useful to probe the eclipse phase for this pulsar at
higher frequencies. We plan to estimate the companion's magnetic 
field near the eclipse boundaries via studying the variation of rotation measure values  \citep{Polzin19,Li19}.

\par
To summarize, in this paper we report a detailed multi-frequency study of the eclipse properties for PSR J1227$-$4853.
In addition to regular eclipses around superior conjunction, the system
also shows evidence of excess dispersion and flux fading around
inferior conjunction.
Simultaneous studies of pulsed and continuum flux densities indicate
flux removal possibly due to the cyclotron absorption rather than 
temporal smearing as the cause of eclipse, both for regular eclipse as well as for
flux fading at inferior conjunction.
\par
We acknowledge support of the Department of Atomic 
Energy, Government of India, under project no. 12-R\&D-TFR-5.02-0700.
The GMRT is run by the National Centre for Radio
Astrophysics of the Tata Institute of Fundamental Research,
India. We acknowledge support of GMRT telescope operators
for observations. We acknowledge discussions with Devojyoti Kanasabanik. 
BWS acknowledges funding from the European Research Council (ERC) under the 
European Union’s Horizon 2020 research and innovation programme (grant agreement No. 694745)

%
\end{document}